\documentclass[%
 reprint,
superscriptaddress,
 amsmath,amssymb,
 aps,
prb,
floatfix,
]{revtex4-2}
\usepackage{graphicx}% Include figure files
\usepackage{dcolumn}% Align table columns on decimal point
\usepackage{bm}% bold math
\usepackage{float}
\usepackage{booktabs}
\usepackage{multirow}
\usepackage{tabularx}
\usepackage[mathlines]{lineno}% Enable numbering of text and display math
\usepackage{braket}
\usepackage[dvipsnames]{xcolor}
\usepackage{amsmath,amssymb,amsfonts} % Typical maths resource package
\usepackage{bbm}
\usepackage{nicefrac}
\usepackage{wasysym}
\usepackage[normalem]{ulem}
\usepackage{hyperref}
\hypersetup{
	colorlinks=true,
	linkcolor=blue,
	filecolor=blue,
	urlcolor=blue,
	citecolor=blue,
}
\usepackage{pifont}
\usepackage{titlesec}
\usepackage{appendix}

\def\kk{{\bm k}}

\begin{document}

%\preprint{APS/123-QED}

\title{Magnetic-Field Tunable M\"{o}bius and Higher-Order Topological Insulators in Three-Dimensional Layered Octagonal Quasicrystals}

\author{Yuxiao Chen}
\author{Zhiming Xu}
\author{Citian Wang}
\affiliation{School of Physics, Peking University, Beijing 100871, China}%

\author{Huaqing Huang}
\email[Corresponding author: ]{huaqing.huang@pku.edu.cn}
\affiliation{School of Physics, Peking University, Beijing 100871, China}%
\affiliation{Collaborative Innovation Center of Quantum Matter, Beijing 100871, China}
\affiliation{Center for High Energy Physics, Peking University, Beijing 100871, China}%

\date{\today}% It is always \today, today,

\begin{abstract}
{We propose that three-dimensional layered octagonal quasicrystals can host magnetic-field-tunable M\"{o}bius insulators and various higher-order topological insulators (HOTIs), enabled by the interplay of quasicrystalline symmetry and magnetic order. By constructing a minimal model based on stacked Ammann-Beenker tilings with magnetic exchange coupling and octagonal warping, we demonstrate that an A-type antiferromagnetic (AFM) configuration yields a topological phase protected by an effective time-reversal symmetry $\mathcal{S}=\mathcal{T}\tau_{1/2}$. Breaking $\mathcal{S}$ via an in-plane magnetic field induced canting of the AFM order while preserving a nonsymmorphic glide symmetry $\mathcal{G}_n=\tau_{1/2}\mathcal{M}_n$ leads to M\"{o}bius-twisted surface states, realizing a M\"{o}bius insulator in an aperiodic 3D system. Furthermore, we show that the quasicrystal with a general magnetic configuration supports multiple HOTI phases characterized by distinct hinge mode configurations that can be switched by rotating the magnetic field. A low-energy effective theory reveals that these transitions are driven by mass kinks between adjacent surfaces. Our work establishes a platform for realizing symmetry-protected topological phases unique to quasicrystals and highlights the tunability of hinge and surface states via magnetic control.}
\end{abstract}

\maketitle

%\tableofcontents
\section{Introduction}
The discovery of time-reversal-symmetric topological insulators (TIs)~\cite{hasan2010colloquium, qi2011topological,moore2010birth, schnyder2008classification} has stimulated extensive interest in novel symmetry-protected topological phases, including topological crystalline insulators (TCIs)~\cite{fu2011topological, hsieh2012topological, tanaka2012experimental, dziawa2012topological, xu2012observation,hsieh2012topological, moore2010birth, schnyder2008classification} and higher-order topological insulators (HOTIs)~\cite{benalcazar2017quantized, PhysRevB.96.245115, song2017d, schindler2018higher, PhysRevLett.119.246401,PhysRevB.97.205136, PhysRevX.8.031070, PhysRevLett.110.046404,huanghq_nwab170}. TCIs generalize the concept of TIs by hosting topological surface states protected by crystalline symmetries rather than time-reversal (TR) symmetry $\mathcal{T}$ alone. While conventional $d$-dimensional TIs and TCIs exhibit gapless states on their
$(d-1)$-dimensional boundaries, HOTIs ($n$-th order TIs with $1 < n \leq d$) are characterized by fully gapped surfaces but support gapless modes on lower ($d-n$)-dimensional boundaries, such as the 1D hinge or 0D corner states of a 3D system.

Recent theoretical progress has extended topological classifications to systems protected by more complex symmetries, particularly due to the intricate magnetic structures. A particularly compelling example is the family of layered magnetic compounds MnBi$_{2n}$Te$_{3n+1}$, where antiferromagnetic (AFM) ordering coexists with topologically nontrivial band structures~\cite{PhysRevLett.122.206401, doi:10.1126/sciadv.aaw5685,Otrokov2019nature, Hu2020Nc,doi:10.1126/sciadv.aax9989, PhysRevX.9.041039, PhysRevX.9.041038, PhysRevX.9.041040}. In these systems, the AFM configuration breaks $\mathcal{T}$ but preserves an effective TR symmetry $\mathcal{S} = \mathcal{T} \tau_{1/2}$, where $\tau_{1/2}$ is a half-lattice translation. This leads to a well-defined $\mathbb{Z}_2$ invariant and an AFM TI phase \cite{PhysRevLett.122.206401}. When an in-plane magnetic field is applied, $\mathcal{S}$ is broken, but if the system retains a nonsymmorphic glide symmetry $\mathcal{G}_n = \tau_{1/2} \mathcal{M}_n$, where $\mathcal{M}_n$ is a mirror reflection perpendicular to the magnetic field, the surface states become M\"{o}bius twisted across the surface Brillouin zone, realizing a so-called M\"{o}bius insulator~\cite{zhang2020Mobius, shiozaki2015z,PhysRevB.93.195413, PhysRevB.99.235105, PhysRevB.102.161117}. Beyond glide mirror symmetry, recent studies have shown that projective translation symmetries can give rise to M\"{o}bius-like topological states in artificial platforms such as phononic crystals \cite{deng2022acoustic, li2022acoustic,PhysRevLett.128.116802,PhysRevB.111.184103}. However, realizations of M\"{o}bius insulators still remain scarce and largely limited to engineered structures.

While most symmetry-protected topological phases have been studied in periodic crystals, their counterparts in quasicrystals remain relatively unexplored. Quasicrystals lack translational periodicity but possess long-range order and noncrystallographic rotational symmetries, such as $\mathcal{C}_5$ or $\mathcal{C}_8$, which are forbidden in periodic crystals. These aperiodic structures can host not only conventional topological phases~\cite{fan2022topological,huanghqQLprl,huanghqQLprb, huanghq2019Comparison,huanghq2019Aperiodic} but also a variety of HOTIs without crystalline analogs~\cite{chen2020higher, PhysRevLett.123.196401, huang2021generic, wang2022effective}. Their unique symmetry properties offer a promising platform for realizing unconventional topological states beyond the standard classification framework.

In this work, we theoretically predict the existence of M\"{o}bius insulator and higher-order topological phases in a 3D layered magnetic quasicrystal with eightfold rotational symmetry. We construct a 3D TI model based on a stacked Ammann-Beenker tiling, incorporating a magnetic exchange interaction $\mathcal{H}_\mathrm{ex}$ and an octagonal warping term $\mathcal{H}_\mathrm{wp}$. We first show that an A-type AFM order along the stacking direction induces an AFM TI state protected by $\mathcal{S}$. Remarkably, we find that an in-plane magnetic field that cants the AFM order along high-symmetry directions stabilizes a glide-symmetry-protected M\"{o}bius phase with twisted surface states. Moreover, we demonstrate that generic magnetic configurations, controlled by the direction of the magnetic field, drive transitions among three distinct HOTI phases (denoted $\alpha$, $\beta$, and $\gamma$), each characterized by a different number and arrangement of hinge modes. By analyzing a low-energy effective model near the center of a pseudo-Brillouin zone, we reveal how the magnetic field induces mass kinks along the intersections of adjacent side surfaces, resulting in field-tunable hinge modes.

\begin{figure}%[tbh]
    \centering
    \includegraphics[width=1\linewidth]{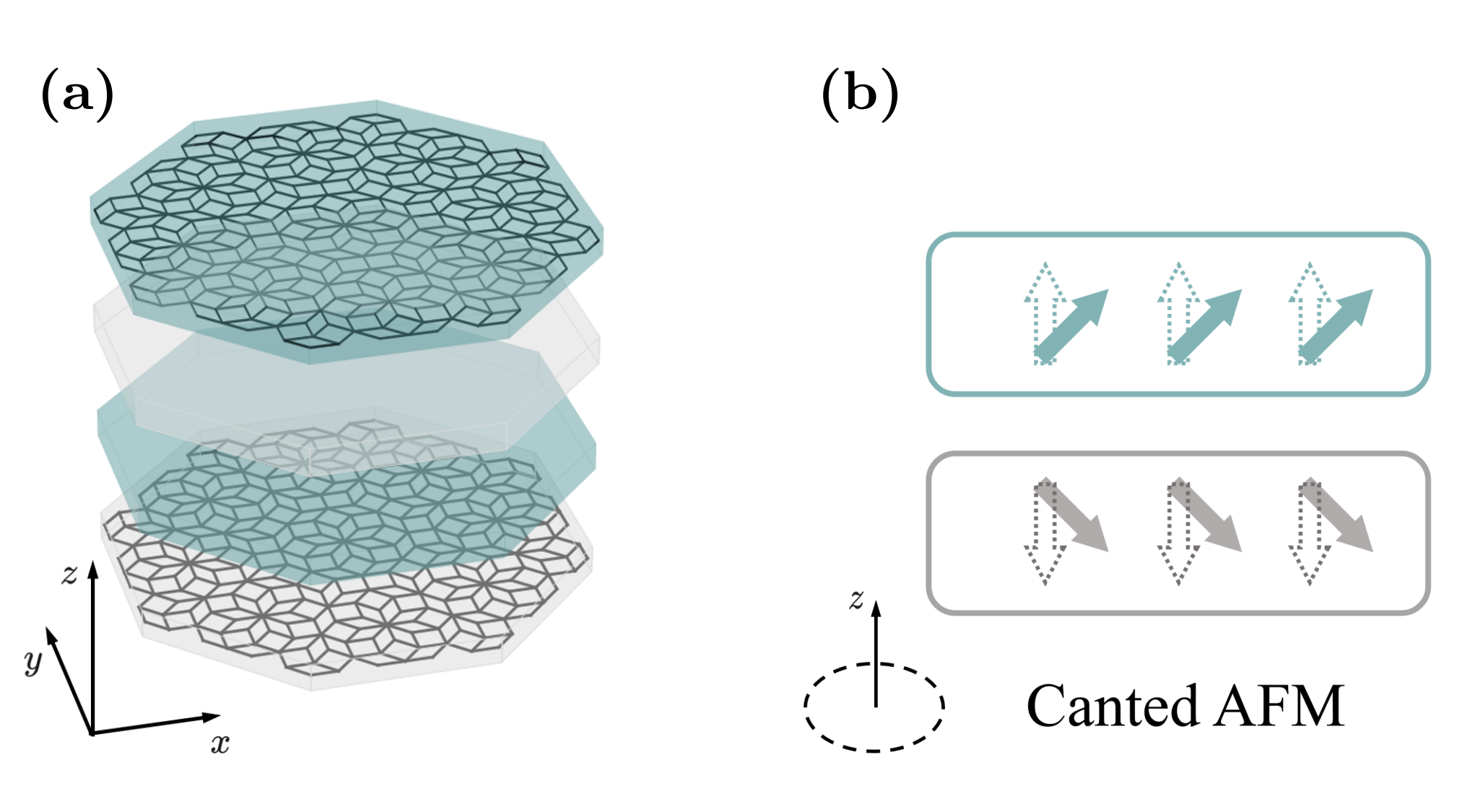}
    \caption{(a) Schematic illustration of a 3D AFM quasicrystal lattice. Each layer in the $x$–$y$ plane forms an Ammann-Beenker tiling (ABT) quasicrystal, exhibiting eightfold rotational symmetry. The ABT quasicrystal layers are stacked along the $z$-axis in an AA stacking configuration. Light blue and gray indicate opposite spin orientations in adjacent layers, representing A-type AFM ordering. (b) When an in-plane magnetic field is applied, the AFM spins cant toward the field direction, breaking certain crystalline symmetries and enabling the realization of M\"{o}bius insulator or various higher-order topological insulators.
    }
    \label{fig:schema}
\end{figure}

\section{Model}

We construct the three-dimensional (3D) quasicrystal model by periodically stacking 2D quasicrystal layers, each characterized by the Ammann-Beenker tiling (ABT), along the out-of-plane ($\hat{\mathbf{z}}$) direction, as illustrated in Fig.~\ref{fig:schema}(a). These layers follow the AA stacking configuration but develop A-type AFM ordering along the $z$ axis. Specifically, we consider a tight-binding  Hamiltonian on the 3D layered AFM quasicrystal lattice, which is given by
\begin{equation}
\begin{aligned}
H= & \sum_{l,j}\mathbf{c}_{lj}^{\dagger}(\mathcal{H}_{0}+\mathcal{H}_\mathrm{ex})\mathbf{c}_{lj}+\sum_{\langle lj,l^\prime k\rangle }\mathbf{c}_{lj}^{\dagger}(\mathcal{H}_{1}+\mathcal{H}_\mathrm{wp})\mathbf{c}_{l^\prime k}, \\
\end{aligned}\label{Ham}
\end{equation}
where
\begin{equation}
\begin{aligned}
\mathcal{H}_{0}=&M_{0}s_{0}\sigma_{3},\\
\mathcal{H}_{1}=& -C_1s_0\sigma_0-M_1s_0\sigma_3+iv(\mathbf{s}\cdot \hat{\mathbf{d}}^{ll^\prime}_{jk})\sigma_1,\\
\mathcal{H}_\mathrm{ex}=&(\mathbf{m}_{l}\cdot \mathbf{s})\sigma_{0},\\
\mathcal{H}_\mathrm{wp}=&-g s_0 \sigma_2 \cos \left(4 \phi_{j k}\right)\delta^{ll'}.
\end{aligned}\label{Ham_comp}
\end{equation}
Here, $l$ denotes the layer index, $j,\ k$ label lattice sites of the quasicrystal within the $x$-$y$ plane, $\mathbf{c}^{\dagger}_{lj}=({c}_{lj, 1}^{\dagger}, {c}_{l j, 2}^{\dagger}, {c}_{lj, 3}^{\dagger}, {c}_{l j, 4}^{\dagger})$ with ${c}_{l j, \alpha}^{\dagger}$ (${c}_{l j, \alpha}$) creates (annihilates) a particle of the $\alpha$-th component at the $j$-th site of the $l$-th layer. $s_{i}$ and $\sigma_{i}$ are the Pauli matrices for the spin and orbital degrees of freedom, respectively. The notation $\langle lj, l'k\rangle$ denotes nearest-neighboring intra- and inter-layer bonds. In Eq. \eqref{Ham} and \eqref{Ham_comp}, $\mathcal{H}_{0}$ represents the on-site energy term with the parameter $M_0$ controlling the band inversion. $\mathcal{H}_{1}$ captures intersite hopping: the $C_1$ and $M_1$ terms describe spin-independent terms, while the third term, proportional to $v$, encodes the spin-orbit coupling (SOC). The directional vector $ \hat{\mathbf{d}}^{ll^\prime}_{jk}=(\cos\phi_{jk}\sin\varphi_{ll^\prime},\sin\phi_{jk}\sin\varphi_{ll^\prime},\cos\varphi_{ll^\prime})$ describes the hopping direction, where $\phi_{jk}$ is the azimuthal angle for the vector from the $k$-th to the $j$-th sites in the same layer with respect to the $x$ axis, and $\varphi_{ll^\prime}$ is the polar angle for interlayer hopping with respect to the $z$ axis. Our model includes only nearest-neighbor intralayer hopping and vertical interlayer hopping between adjacent layers in the AA stacking configuration (i.e., $l^\prime=l$ or $l\pm1$).

In the absence of an external magnetic field, our system exhibits interlayer AFM ordering. Upon application of an external field $\mathbf{B}$, the AFM coupling can be successively tuned into canted AFM, canted ferromagnetic (FM), or even uniform FM states. To model this layer-dependent magnetic order, we introduce an exchange-coupling term $\mathcal{H}_\mathrm{ex}$. The average exchange field acting on the local spin in layer $l$ is characterized by
\begin{equation}
    \mathbf{m}_{l} = m(\cos\phi\sin\theta,\ \sin\phi\sin\theta,\ \pm\cos\theta).\label{m_l}
\end{equation}
Here, the sign of the $z$-component alternates between odd and even layers to describe the (canted) AFM states. For the canted FM states, the $z$-component takes a uniform sign across all layers.
In addition, we include the term $\mathcal{H}_\mathrm{wp}$ to describe the octagonal warping effect, which may arise from magnetic systems with higher-order SOC \cite{wan2024topological,zhang2020Mobius}. It is also known as a TR-breaking Wilson mass term, which is responsible for higher-order topology \cite{PhysRevResearch.2.012067, PhysRevLett.124.036803,PhysRevLett.123.196401,PhysRevB.102.241102,PhysRevB.108.195306}. Unless otherwise specified, all calculations in this study are performed on an ABT quasicrystal lattice comprising 1273 sites in the $x$–$y$ plane.

\section{Results and discussion}

\begin{table}%[!ht]
\centering
\caption{Topological phases of the model [Eq.~\eqref{Ham} and \eqref{Ham_comp}] in stacking ABT quasicrystals under different conditions.}
\label{tab:topo_class}
\begin{tabular*}{\linewidth}{@{\extracolsep{\fill}}cccc}
\toprule[1.5pt]
\addlinespace[5pt]
\textbf{$\mathcal{H}_{\mathrm{wp}}$} & \textbf{$\mathcal{H}_{\mathrm{ex}}$} & \textbf{Symmetry}& \textbf{Topology} \\
\addlinespace[3pt]
\midrule[1pt]
\ding{55} & \ding{55} & $\mathcal{T}$ & TI \\
\addlinespace[3pt]
\ding{55} & AFM-$z$ & $\mathcal{T}\tau_{1/2}$ & AFM TI \\

\addlinespace[3pt]
\ding{51} & \ding{55} & $\mathcal{C}_8\mathcal{T}$ & HOTI \\

\addlinespace[3pt]
\ding{51} & AFM-$z$ & $\mathcal{C}_8\mathcal{T}\tau_{1/2}$ & HOTI \\

\addlinespace[3pt]
\hline
\addlinespace[3pt]
\ding{55} & canted AFM along $\hat{\mathbf{x}}$&$\mathcal{G}_x$ & M\"{o}bius insulator \\
%\ding{55}    & canted FM & $\mathcal{P}$ & HOTI $\beta$ \\
%\ding{55}    & FM & $\mathcal{P}$ & HOTI $\alpha$ \\
%\multirow{3}{*}{\ding{51}}
\ding{51} & canted AFM&\ding{55} & HOTI $\beta$,\ $\gamma$ \\
\ding{51}    & canted FM & \ding{55} & HOTI $\beta$,\ $\gamma$ \\
\ding{51}    & FM & \ding{55} & HOTI $\alpha$ \\
\bottomrule[1.5pt]
\end{tabular*}
\end{table}

\begin{figure}%[tbh]
    \centering
    \includegraphics[width=1\linewidth]{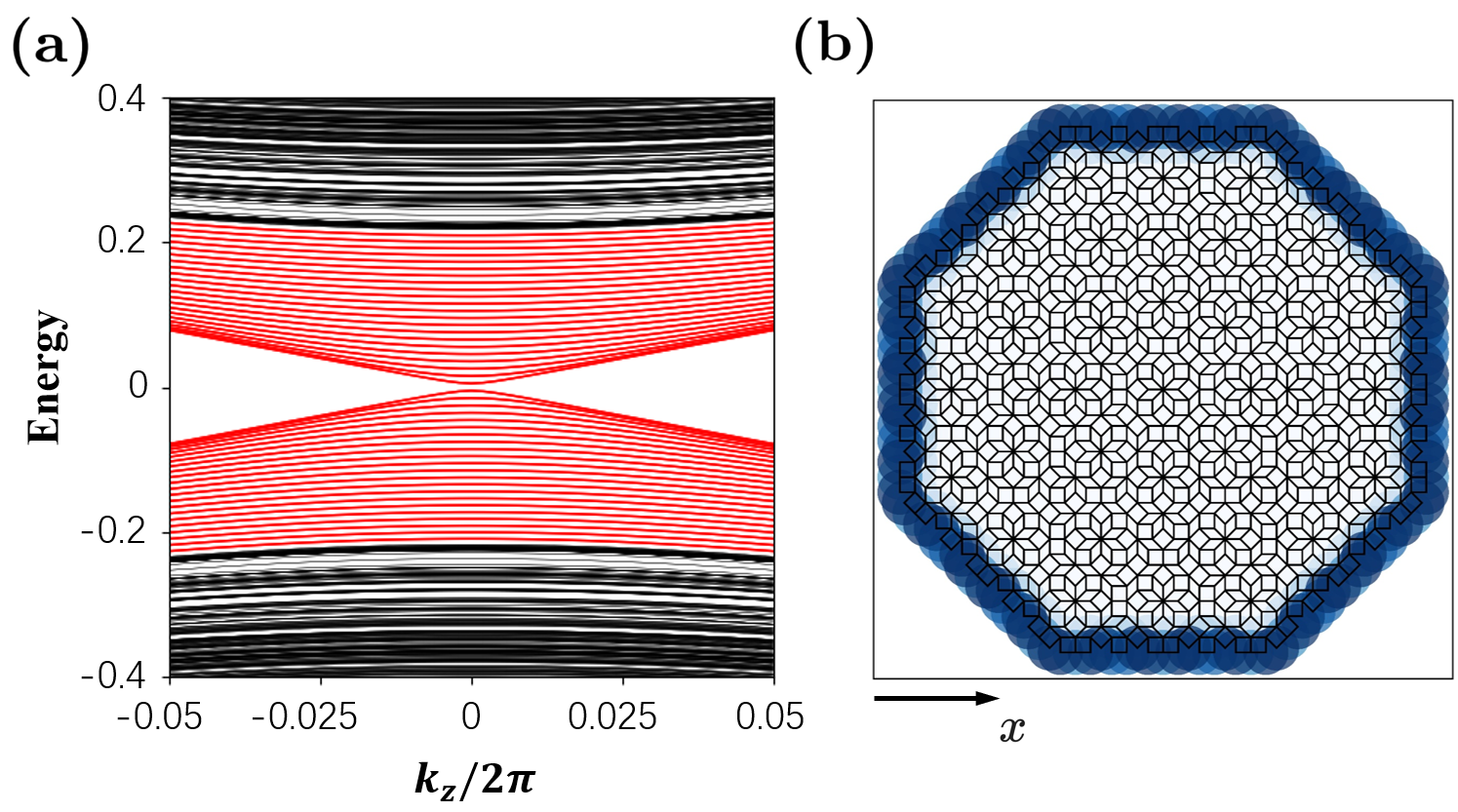}
    \caption{Energy spectrum and spatial distribution of the surface state for the AFM TI phase. (a) Band structure along the $k_z$ direction for the AFM TI in an octagonal prime geometry. The red lines represent surface states. (b) (Top view along the $z$ direction) Wavefunction amplitude distribution of the surface state corresponding to $k_z=0$. The calculation is performed using Eq.~\eqref{Ham} and \eqref{Ham_comp} with the parameters $C_1=0.5,\ M_1=0.5,\ v=0.5,\ M_0 = -2$, $g = 0$, $m = 0.4$, $\theta = 0$, and $\phi = 0$.
    }
    \label{fig:AFM-TI}
\end{figure}

\subsection{AFM TI and HOTI}% (AFM TI)}
Before exploring the magnetically tunable M\"{o}bius insulator and HOTIs, let us first analyze two limit cases of our model Hamiltonian. We can identify the basis states as $|\pm,\uparrow(\downarrow)\rangle$ with opposite parities and both spins, which are the linear combinations of $p_z$ orbitals \cite{PhysRevB.82.045122}. In the absence of $\mathcal{H}_{\mathrm{ex}}$ and $\mathcal{H}_{\mathrm{wp}}$, the system possesses both spatial inversion symmetry $\mathcal{P}$ and TR symmetry $\mathcal{T}$. By tuning $M_0$ to the region of band inversion, Eq.~\eqref{Ham} behaves as a conventional 3D TI \cite{PhysRevB.82.045122}. Upon introducing the AFM order along the $z$ direction (i.e., $\mathbf{m}_{l}=(0,\ 0,\ \pm m)$), the TR symmetry $\mathcal{T}$ is broken; however, a combined symmetry $\mathcal{S}=\mathcal{T}\tau_{1/2}$ is preserved, where $\tau_{1/2}$ is the half translation operator connecting nearest spin-up and -down quasicrystal layers. The system becomes an AFM TI ~\cite{PhysRevB.81.245209, PhysRevLett.122.206401}. As illustrated in Fig.~\ref{fig:AFM-TI}, the system hosts surface states on all open side facets in an octagonal prism geometry.

In the alternative limit where $\mathcal{H}_{\mathrm{ex}}=0$ and a finite warping term $\mathcal{H}_{\mathrm{wp}}$ is introduced, both the TR symmetry $\mathcal{T}$ and 8-fold rotational symmetry $\mathcal{C}_8=e^{-\frac{i\pi}{8}s_3\sigma_0}\mathcal{R}_8$ are broken, but the combined symmetry $\mathcal{T}\mathcal{C}_8$ is preserved. Consequently, the $\mathcal{T}$-protected surface states are gapped because of the octagonal warping-generated mass terms. However, the combined symmetry $\mathcal{TC}_8$ requires the mass terms to switch sign alternating on the side surfaces of the octagonal prism geometry. This enables the emergence of eight hinge modes on the octagonal prism, implying the existence of a HOTI state. Prior works \cite{mao2023higher, PhysRevB.108.195306} have shown that such a warping term induces a higher-order topological phase with hinge states on the octagonal prism edges.
In the presence of both the AFM type $\mathcal{H}_\mathrm{ex}$ and $\mathcal{H}_\mathrm{wp}$, the system is still a HOTI but protected by the combined symmetry $\mathcal{C}_8\mathcal{T}\tau_{1/2}$. This is because $\mathcal{H}_\mathrm{wp}$ remains unchanged under $\tau_{1/2}$ operation, while $\mathcal{H}_\mathrm{ex}$ with $\mathbf{m}_{l}=(0,\ 0,\ \pm m)$ respects $\mathcal{C}_8$ symmetry.

In the following, we will show that more intriguing topological phases and transitions can be realized by applying a magnetic field. For a clear illustration, Table~\ref{tab:topo_class} summarizes various topological phases discovered in the 3D layered magnetic quasicrystal based on the model in Eq.~\eqref{Ham} and \eqref{Ham_comp} within different parameter regimes.

\begin{figure}%[tbh]
    \centering
    \includegraphics[width=1\linewidth]{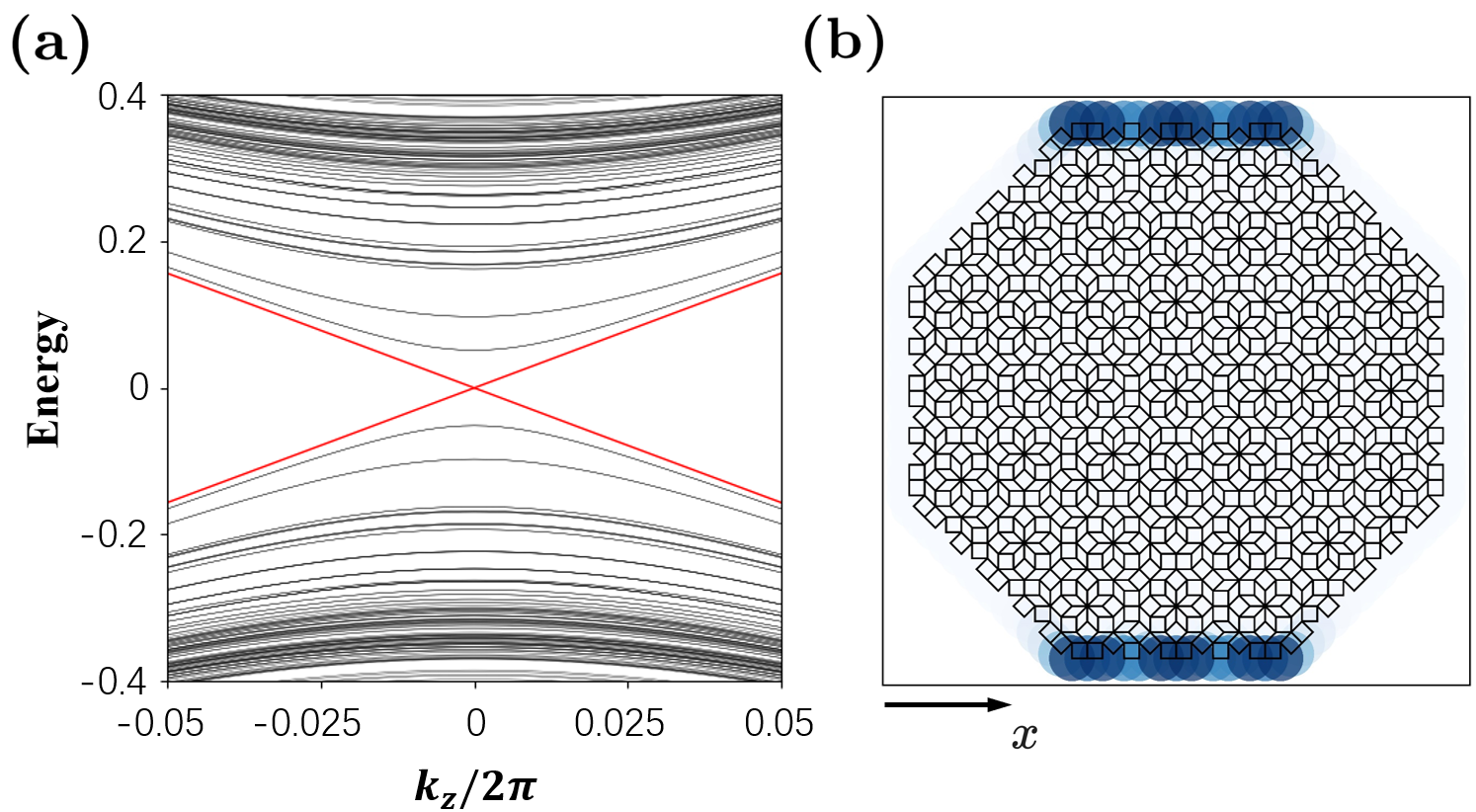}
    \caption{Energy spectrum and spatial distribution of the surface state for the M\"{o}bius insulator phase. (a) Band structure along $k_z$ for an octagonal prism geometry with periodic boundary conditions along the $z$ direction. The red lines represent gapless surface states. (b) (Top view along the $z$ direction) Wavefunction amplitude distribution of the surface states corresponding to the Dirac cone at $k_z=0$, which exist only on the front and rear side facets. Since $\mathbf{B}=B\hat{\mathbf{x}}$, the AFM cants along the field direction, which is modeled by Eq.~\eqref{m_l} with parameters $\theta = 0.3\pi$ and $\phi = 0$. Other parameters are the same as Fig.~\ref{fig:AFM-TI}.} %In these calculations, $C_1=0.5,\ M_1=0.5,\ v=0.5,\ M_0 = -2$, $g = 0$, $m = 0.4$, $\theta = 0.3\pi$, and $\phi = 0$.}
    \label{fig:Mobius}
\end{figure}

\begin{figure}[htb]
    \centering
    \includegraphics[width=1\linewidth]{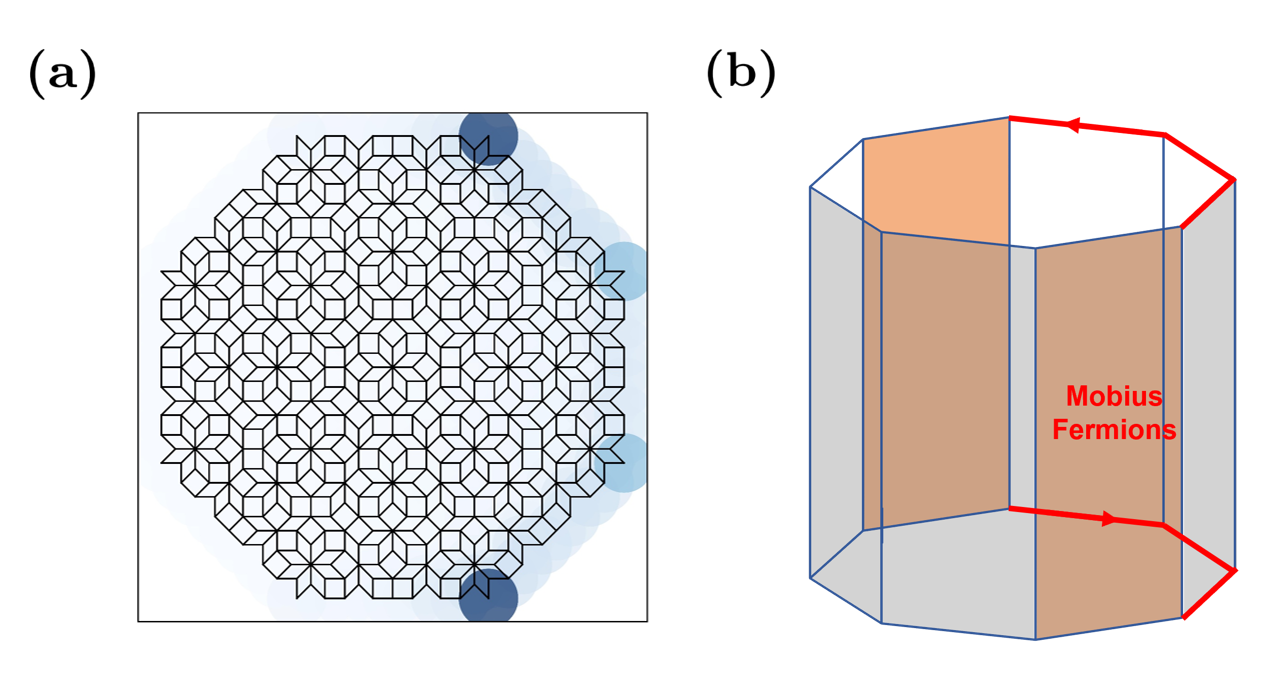}
    \caption{Higher-order M\"{o}bius insulator in a finite octagonal prism geometry with open boundary conditions along the stacking direction. The system is composed of 5 layers of quasicrystal lattices, each comprising 577 sites. (a) (Top view along the $z$ direction) Wavefunction amplitude distribution of the hinge states. (b) Schematic illustration of the higher-order M\"{o}bius insulator. The red lines with arrows represent hinge modes connecting M\"{o}bius surface states on the front and rear side surfaces.}
    \label{fig:HOMI}
\end{figure}

\subsection{M\"{o}bius insulator in canted AFM phases}
We now analyze a representative topological configuration featuring M\"{o}bius-twisted surface states under the condition $\mathcal{H}_\mathrm{wp}=0$. By applying an external in-plane magnetic field $\mathbf{B}$ to the system, the AFM order develops a transverse canted structure, as shown in Fig.~\ref{fig:schema}(b). When the magnetic field is applied along the $\hat{\mathbf{x}}$ direction (i.e., $\mathbf{B}=B\hat{\mathbf{x}}$), the quasicrystal Hamiltonian respects a global glide-reflection symmetry $\mathcal{G}_x = \mathcal{\tau}_{(00\frac{1}{2})}\mathcal{M}_x$. This $\mathcal{G}_x$ symmetry protects the gapless surface states on the glide-symmetric front and rear faces of the octagonal prism (Fig.~\ref{fig:Mobius}), whereas other surfaces break the symmetry and become gapped. To understand this protection mechanism, we first turn to the symmetry-group analysis of the quasicrystal approximants, which share the same glide-reflection symmetry and HOTI as the quasicrystals. In the $k_x = 0, \pi$ planes in the Brillouin zone of a quasicrystal approximant, the wave-vector group satisfies $\mathcal{G}_x$ symmetry, allowing surface states to be labeled by distinct eigenvalues $g_x = \pm i e^{i k_z / 2}$. Crucially, half-unit cell translation in $\mathcal{G}_x$ endows the eigenvalue $g_x(k_z)$ with a $4\pi$ periodicity in momentum space. This periodicity $4\pi$ forces the phase of $g_x$ to traverse a noncontractible path, resulting in a M\"{o}bius strip topology for the eigenvector bundle associated with the surface states. Since one can approach the true quasicrystal limit asymptotically by enlarging the quasicrystal approximants \cite{jagannathan2024propertiesammannbeenkertilingsquare}, and given that the above reasoning holds rigorously for approximants of any finite size, the M\"{o}bius state persists in real quasicrystalline system. This generates a locally robust Dirac point, whose M\"{o}bius character is confirmed by the energy spectral analysis and surface-state wavefunction distribution, as shown in Fig.~\ref{fig:Mobius}.

Remarkably, although M\"{o}bius-twisted surface states reside only on two opposite side facets where glide symmetry is preserved and surface gaps open on all other facets, we find the presence of hinge-localized modes that connect these gapless surfaces in a finite octagonal prism geometry with open boundary conditions along the stacking direction (see Fig.~\ref{fig:HOMI}). This reveals the emergence of higher-order topological character in the magnetic-field controllable M\"{o}bius phase, when the warping term $\mathcal{H}_\mathrm{wp}$ is absent. The resulting phase---a higher-order M\"{o}bius insulator---features the coexistence of M\"{o}bius-twisted surface states and 1D hinge modes, and may be viewed as a quasicrystalline analog of that proposed in MnBi$_{2n}$Te$_{3n+1}$, where both glide symmetry $\mathcal{G}_x$ and inversion symmetry $\mathcal{P}$ play a crucial role \cite{zhang2020Mobius}.

\begin{figure}%[tbh]
    \centering
    \includegraphics[width=1\linewidth]{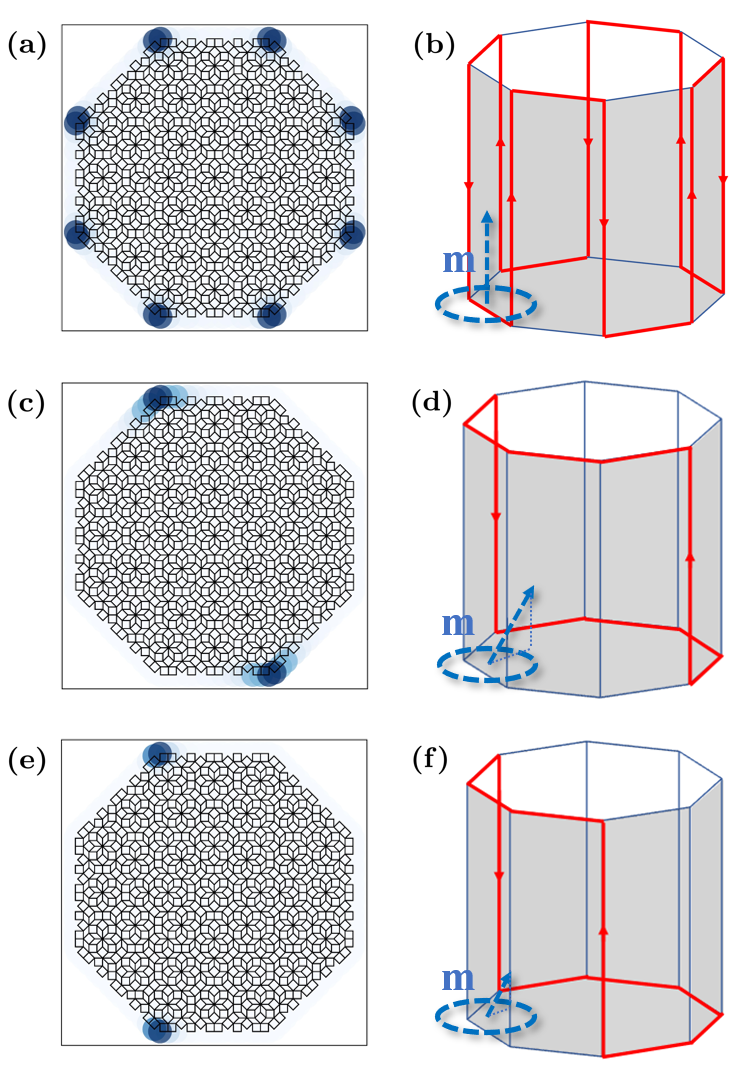}
    \caption{Magnetic-field tunable HOTI phases. (a),(c),(e) Wavefunction amplitude distribution of HOTI $\alpha$, $\beta$, and $\gamma$ phase, respectively. (b),(d),(f) Schematic illustration of the chiral current in three-dimensional space for the hinge modes shown in the left figure. The arrow in the lower-left corner indicates the direction of the magnetization $\mathbf{m}$. The parameters used for (a) $g = 0.25$, $\theta = 0$, and $\phi = 0$; (c) $g = 0$, $\theta = 0.3\pi$, and $\phi = \pi/8$; and (e) $g = 0.25$, $\theta = 0.3\pi$, and $\phi = 0$.     Other parameters are the same for the three cases, which are $C_1=0.5,\ M_1=0.5,\ v=0.5,\ M_0 = -2$, and $m = 0.4$.}
    \label{fig:3HOTI}
\end{figure}

\subsection{Magnetically tunable transitions between HOTIs}
When $\mathbf{B}$ deviates from the high-symmetric $\hat{\mathbf{x}}$ axis, a variety of new magnetic configurations can arise, typically breaking the glide symmetry $\mathcal{G}_x$ and gapping the 2D M\"{o}bius surface states. Despite this, 1D hinge modes remain robust along the hinges of the octagonal prism, signaling the persistence of higher-order topology. In the absence of the warping term $\mathcal{H}_{\mathrm{wp}}$, the HOTI in the canted AFM quasicrystal is protected by inversion symmetry $\mathcal{P}$. However, since $\mathcal{H}_{\mathrm{wp}}$ is odd under $\mathcal{P}$ operation, its presence breaks the inversion symmetry and can potentially destabilize the HOTI phase. Nevertheless, we find that the HOTI remains stable for a wide range of magnetic configurations when $\mathcal{H}_\mathrm{wp}$ is relatively weak. Notably, we identify three distinct classes of HOTI phases (labeled $\alpha, \beta$, and $\gamma$) that emerge under various canted AFM and persist even in FM regime, as the AFM order can eventually evolve to FM by gradually increasing the magnetic field strength. These higher-order topological phases can be continuously tuned by varying the orientation of the applied magnetic field, as discussed next.

When the system is in a FM order, the magnetic exchange coupling [Eq.~\eqref{Ham_comp} and \eqref{m_l}] simplifies to $\mathbf{m}_{2n+1}=\mathbf{m}_{2n}= m_z\hat{\mathbf{z}}+ m_{\parallel}\hat{\mathbf{n}}$, indicating a uniform FM alignment across all layers. In the special case of FM along the $\hat{\mathbf{z}}$ axis (i.e., $m_{\parallel}=0$), this configuration yields eight hinge modes. The spatial distributions and propagation directions of these hinge states are illustrated in Fig.~\ref{fig:3HOTI}(a) and~\ref{fig:3HOTI}(b), with half propagating upward and half downward along the vertical hinges. The combination of the out-of-plane FM and the octagonal warping term opens a gap in the surface Hamiltonian on all side surfaces. The resulting masses on adjacent side surfaces acquire opposite signs, leading to the formation of mass domain walls at the hinges. According to the Jackiw-Rebbi mechanism \cite{jackiw1976solitons}, these mass domain walls support zero-energy modes, giving rise to robust 1D hinge modes. This configuration realizes a HOTI, denoted as the HOTI $\alpha$ phase.

When the external magnetic field acquires an in-plane component that deviates from the high-symmetry $\hat{\mathbf{z}}$ axis, the system enters a canted FM state characterized by a nonzero in-plane magnetization component, $m_{\parallel}\neq 0$. In this regime, the topological character of the system is determined by the competition between the magnetic exchange coupling $\mathcal{H}_\mathrm{ex}$ and the octagonal warping potential $\mathcal{H}_\mathrm{wp}$, quantified by the ratio $|\mathbf{m}|/g$, where $|\mathbf{m}|$ is the strength of in-plane magnetic exchange coupling and $g$ is the amplitude of warping term in Eq.~\eqref{Ham_comp}. For sufficiently strong magnetization, the magnetic exchange interaction dominates, effectively suppressing the octagonal warping effects. This leads to the formation of only two surface domains with opposite Dirac mass terms, separated by the longest diagonal of the octagon---perpendicular to the in-plane magnetization direction \cite{PhysRevLett.127.066801,PhysRevLett.124.166804,PhysRevLett.108.126807, PhysRevLett.110.046404}.
As a result, two hinge states predominantly localize at the endpoints of this diagonal, %where two hinge states are separated by four faces of the octagonal prism
as shown in Fig.~\ref{fig:3HOTI}(c) and \ref{fig:3HOTI}(d). This regime defines the HOTI $\beta$ phase.
In contrast, when the magnetic exchange coupling is weaker relative to the warping potential and unable to fully suppress the alternating mass domain induced by $\mathcal{H}_\mathrm{wp}$, the system transitions into the HOTI $\gamma$ phase. In this phase, two hinge modes no longer reside at the endpoints of the longest diagonal but instead appear along different edges of the octagonal prism, as illustrated in Figs.~\ref{fig:3HOTI}(e) and \ref{fig:3HOTI}(f).

Remarkably, our numerical results reveal a layer-parity effect in the spatial localization of hinge modes. When the total number of layers is odd, the spatial pattern of both the HOTI $\beta$ and $\gamma$ phases remains unchanged as the magnetic configuration is varied from a canted FM to a canted AFM type, provided all other parameters are fixed. However, for even numbers of layers, %which does not preserve inversion symmetry for AFM order,
the hinge modes exhibit a distinct “inversion-asymmetric” distribution, as previously discussed in Ref.~\cite{PhysRevResearch.2.043274}. These findings indicate that the HOTI $\beta$ and $\gamma$ phases can be stabilized in both canted FM and canted AFM configurations, with the specific phase determined primarily by the ratio $|\mathbf{m}|/g$.

\begin{figure}%[tbh]
    \centering
    \includegraphics[width=1\linewidth]{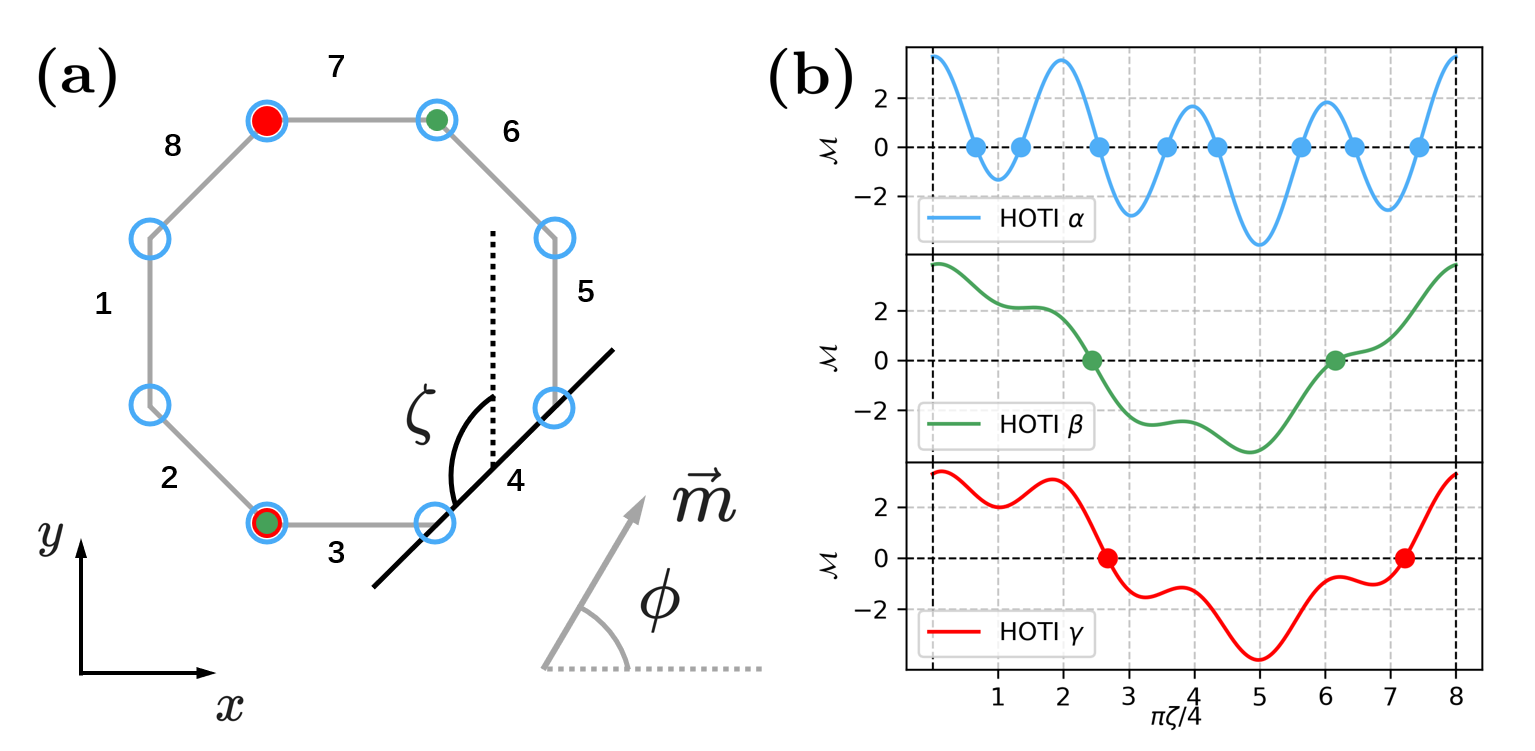}
    \caption{(a) Schematic illustration of the hinge mode configuration of HOTI $\alpha,\beta$, and $\gamma$. Blue, green, and red dots indicate the real-space distribution of hinge modes under the HOTI $\alpha$, $\beta$, and $\gamma$ phases. The lateral facets are marked from 1 to 8. $\vec{m}$ denotes the in-plane magnetization direction. (b) Mass term $\mathcal{M}$ in Eq.~\eqref{mass_term} as a function of the polar angle $\zeta$ for three distinct types of HOTI phases. The integer ticks on the $x$-axis are related to lateral facets labelled in (a).}
    \label{fig:mass}
\end{figure}

\subsection{Effective surface theory of the octagonal magnetic quasicrystal}
To gain physical insight into the transition between these HOTI phases, we adopt the effective $\kk$-space formulation developed in Ref.~\cite{wang2022effective} to construct an approximate low-energy effective Hamiltonian for the quasicrystal. Given the periodicity along the $z$ direction, $k_z$ remains a good quantum number. In the long-wavelength limit, we average over intersite hoppings in the quasicrystalline lattice and derive an effective continuum model near the center of the pseudo-Brillouin zone of the 3D layered octagonal quasicrystal under a canted FM order. The resulting Hamiltonian takes the form
\begin{equation}
\begin{aligned}
\mathcal{H}_\mathrm{eff}(\kk)=& [\widetilde{C}+\widetilde{C}_{1}k_{z}^{2}+\widetilde{C}_{2}(k_{x}^{2}+k_{y}^{2})]s_{0}\sigma_{0} \\
&+ [\widetilde{M}+\widetilde{M}_{1}k_{z}^{2}+\widetilde{M}_{2}(k_{x}^{2}+k_{y}^{2})]s_{0}\sigma_{3}\\
 & -v_{1}k_{z}s_{3}\sigma_{1}-v_{2}(k_{x}s_{1}+k_{y}s_{2})\sigma_{1} \\
 & +\widetilde{g}[k_{x}^{2}k_{y}^{2}-\frac{1}{6}(k_{x}^{4}+k_{y}^{4} )]s_{0}\sigma_{2} \\
 & +(m_zs_3+m_\parallel\cos\phi s_1+m_\parallel\sin \phi s_2)\sigma_{0},\\
\end{aligned}\label{Hkp_Gamma}
\end{equation}
where $\widetilde{C}_i$, $\widetilde{M}_{i}$, $v_i$, and $\widetilde{g}$ are effective model parameters (see Appendix \ref{app:A}). Here, $k_x$ and $k_y$ denote in-plane momenta around the $\Gamma$ point in the reciprocal space, and $\phi$ is the angle between the in-plane magnetization component $\vec{m}_\parallel$ and the $x$-axis [see Fig.~\ref{fig:mass}(a)]. The last two lines originate from the octagonal warping term $\mathcal{H}_\mathrm{wp}$ and the magnetic exchange coupling $\mathcal{H}_\mathrm{ex}$, respectively.

To analyze the side surface state, we consider a lateral facet of the octagonal prism with normal vector $\vec{e}_1=\cos \zeta \hat{e}_x+\sin \zeta \hat{e}_y$, where $\zeta$ denotes the polar angle relative to the $y$-axis (see Appendix \ref{app:B}). We perform a rotation of the momentum components via $ k_x=k_1 \cos \zeta-k_2 \sin \zeta, k_y=k_1 \sin \zeta+k_2 \cos \zeta$ and replace $k_1\rightarrow-i\partial_{x_1}$ for the momentum perpendicular to the surface. Solving the effective Hamiltonian on the surface---neglecting the in-plane magnetization and octagonal warping terms---yields gapless Dirac surface states with the projected surface Hamiltonian
\begin{equation}
\mathcal{H}^0_{\mathrm{surf}}=-\bar{v}_{1}k_{z}\sigma_{1}-(\bar{v}_{2}k_{2}+\bar{m})\sigma_{3},
\end{equation}
where $\bar{v}_{1},\ \bar{v}_{2}$, and $\bar{m}$ are parameters independent of $k_z,\ k_1$ or $k_2$. This describes a Dirac cone centered at $k_{2}=-\bar{m}/\bar{v}_{2}$. When the in-plane magnetic exchange coupling and warping terms are included, the Dirac surface states are gapped by mass terms
\begin{equation}
    \mathcal{M}=\mathcal{M}_0^{\mathrm{ex}}\cos (\zeta-\phi)\sigma_2+\mathcal{M}^{\mathrm{wp}}_0\cos (4\zeta)\sigma_2\label{mass_term},
\end{equation}
where $\mathcal{M}_0^{\mathrm{ex}}\propto m_{\parallel}$ and $\mathcal{M}^{\mathrm{wp}}_0\propto g$ are mainly determined by the strength of in-plane magnetic exchange coupling and the octagonal warping, respectively
(Details are discussed in Appendix \ref{app:B}).

The transition between HOTIs with distinct hinge mode configurations can be understood by inspecting the spatial variation of the mass term $\mathcal{M}$ across the lateral facets. Figure~\ref{fig:mass}(b) illustrates the evoluation of $\mathcal{M}$ as a function of the polar angle $\zeta$ for representative configurations. When $\mathcal{M}^{\mathrm{wp}}_0\gg\mathcal{M}_0^{\mathrm{ex}}$, the mass term alternates sign between adjacent facets, leading to eight domain walls and corresponding hinge modes---characteristic of the HOTI $\alpha$ phase. Conversely, in the regime $\mathcal{M}^{\mathrm{wp}}_0\ll\mathcal{M}_0^{\mathrm{ex}}$, the mass term switches sign only at two opposite hinges connected by a diagonal of the octagon, resulting in two hinge modes and the HOTI $\beta$ phase. At intermedia values where $\mathcal{M}_0^{\mathrm{ex}}\sim\mathcal{M}^{\mathrm{wp}}_0$, the mass pattern reflects a competition between the magnetic exchange coupling and the octagonal warping terms. This competition leads to a flexible and asymmetric distribution of hinge modes, corresponding to the HOTI $\gamma$ phase. While our effective theory also allows for configurations with four or six hinge modes, such intermediate HOTI states were not observed in our numerical simulations---possibly due to the limited parameter range and system size explored.

\section{Conclusion}
We have demonstrated the emergence of magnetic-field-tunable M\"{o}bius and multiple HOTI phases in a 3D layered magnetic quasicrystal with octagonal rotational symmetry. Specifically, we proposed a 3D TI model incorporating magnetic exchange coupling and octagonal warping in a stack of Ammann-Beenker tilings. Within this framework, we identified a higher-order M\"{o}bius topological phase protected by glide mirror symmetry in a canted AFM configuration---representing the first theoretical proposal of M\"{o}bius topology in a quasicrystalline system.

Under an external magnetic field, the interplay between magnetic ordering and octagonal warping gives rise to a rich phase diagram featuring multiple HOTI phases. These phases, characterized by distinct hinge mode configurations, can be tuned by varying both the magnitude and orientation of the external magnetic field. Through an effective surface Hamiltonian analysis, we showed that the exchange interaction $\mathcal{H}_\mathrm{ex}$ and the warping term $\mathcal{H}_\mathrm{wp}$ generate surface mass terms, and that the sign alternation of these masses across adjacent facets---controlled by the magnetic field---plays a crucial role in stabilizing different HOTI phases. This mechanism provides a unified theoretical framework for understanding these field-driven transitions in quasicrystalline systems.

Our results establish a theoretical foundation for exploring higher-order and M\"{o}bius topology in quasicrystals and emphasize the versatility of magnetic and structural tuning in controlling topological phases. While our study is theoretical, it offers guiding principles for future experimental realizations and points to the potential of quasicrystals as a platform for novel electronic and topological functionalities. These findings contribute to the broader understanding of symmetry-protected topological phases in non-periodic systems and open new avenues for quasicrystal-based topological materials.

\begin{acknowledgments}
This work is supported by the National Key R\&D Program of China (Grant No. 2021YFA1401600), the National Natural Science Foundation of China (Grants No. 12474056), and the 2022 basic discipline top-notch students training program 2.0 research project (Grant No. 20222005). The computational resources are supported by the high-performance computing platform of Peking University. We thank Zhi-Cheng Yang and Jun Leng for valuable discussions.
\end{acknowledgments}

\appendix

\begin{widetext}
\section{Effective k-Space Hamiltonian}
\label{app:A}
To construct an effective $k\cdot p$ Hamiltonian around the center of the pseudo-Brillouin zone for the 3D qusicrystal, we follow the procedure introduced in Ref.~\cite{wang2022effective}. Specifically, we first derive an effective model for the Hamiltonian without exchange coupling and octagonal warping, and obtain the surface Hamiltonian. Then, we treat the $\mathcal{H}_\mathrm{ex}$ and $\mathcal{H}_\mathrm{wp}$ as perturbations, which are expected to generate mass terms for the surface Hamiltonian. To do so, we begin by regrouping Eq.~\eqref{Ham} as
$H=H_\perp+H_\parallel+H_\mathrm{ex}+H_\mathrm{wp}$
where
\begin{eqnarray}
    H_\perp&=& \sum_{l,j}\mathbf{c}_{lj}^\dagger\mathcal{H}_0\mathbf{c}_{lj}+\sum_{\langle l,l^\prime\rangle,j}\mathbf{c}_{lj}^\dagger\mathcal{H}_{1z}\mathbf{c}_{lk} \\
    &=&\sum_{l,j}\mathbf{c}_{lj}^\dagger M_{0}s_{0}\sigma_{3}\mathbf{c}_{lj} +\sum_{\langle l,l^\prime \rangle,j}\mathbf{c}_{lj}^\dagger\left(-C_1s_0\sigma_0-M_1s_0\sigma_3+iv(\mathbf{s}\cdot \hat{\mathbf{d}}^{ll^\prime}_{jj})\sigma_1\right)\mathbf{c}_{l^\prime j},\\
    H_\parallel&=&\sum_{l,\langle j,k\rangle}\mathbf{c}_{lj}^\dagger\mathcal{H}_{1xy}\mathbf{c}_{lk}=\sum_{l,\langle j,k\rangle}\mathbf{c}_{lj}^\dagger\left(-C_1s_0\sigma_0-M_1s_0\sigma_3+iv(\mathbf{s}\cdot \hat{\mathbf{d}}^{ll}_{jk})\sigma_1\right)\mathbf{c}_{lk},\\H_\mathrm{ex}&=&\sum_{l,j}\mathbf{c}_{lj}^\dagger\mathcal{H}_\mathrm{ex}\mathbf{c}_{lj}=\sum_{l,j}\mathbf{c}_{lj}^\dagger(\mathbf{m}_{l}\cdot \mathbf{s})\sigma_{0}\mathbf{c}_{lj},\\
    H_\mathrm{wp}&=&\sum_{\langle lj,l'k\rangle}\mathbf{c}_{lj}^\dagger\mathcal{H}_\mathrm{wp}\mathbf{c}_{l'k}=\sum_{\langle lj,l'k\rangle}\mathbf{c}_{lj}^\dagger(-g s_0 \sigma_2 \cos \left(4 \alpha_{j k}\right))\mathbf{c}_{l'k}.
\end{eqnarray}

Since the lattice is periodic along the $z$ direction, $k_z$ is a good quantum number. We perform a Fourier transform along the $z$-direction, which replaces $\mathcal{H}_{\perp}$ with
\begin{equation}
    \mathcal{H}_{\perp}^{k_{z}}=\mathcal{H}_{0}+e^{ ik_{z} }\mathcal{H}_{1z}+e^{ -ik_{z} }\mathcal{H}_{1z}^{\dagger}.
\end{equation}
The total Hamiltonian $H$ then becomes
\begin{equation}
\begin{aligned}
H  =\sum_{k_{z}}\left[ \sum_{j}\mathbf{c}_{k_{z}j}^{\dagger}(\mathcal{H}^{k_{z}}_{\perp}+\mathcal{H}_\mathrm{ex})\vec{c}_{k_{z}j}+ \sum_{\langle j,\ k\rangle }\vec{c}_{k_{z}j}^{\dagger}(\mathcal{H}_{\parallel} +\mathcal{H}_\mathrm{wp})\mathbf{c}_{k_{z}k}\right]
\end{aligned}
\end{equation}
For the in-plane part in the quasicrystal lattice, we apply the JZ Fourier transformation~\cite{wang2022effective, JIANG2014428} to obtain the expression in the 4D periodic reciprocal space for the Ammann-Beenker tiling. In this formalism, a function $g(\mathbf{r})$ defined in the 2D quasicrystalline lattice can be expanded as
\begin{equation}
g(\mathbf{r})=\sum_{\boldsymbol{\Pi}} \hat{g}(\boldsymbol{\Pi}) e^{i\left[(\mathcal{S} \cdot \boldsymbol{\Pi})^T \cdot \mathbf{r}\right]},
\end{equation}
where $\mathbf{r}$ represents lattice sites in the 2D physical space spanned by $\mathbf{e}_{x},\ \mathbf{e}_{y}$, $\Pi$ are 4D reciprocal space vectors in the 4D reciprocal space, and $\mathcal{S}$ is the projection matrix that maps the 4D reciprocal hyperspace to the 2D physical space. Using this expansion, both on-site and hopping terms can be simplified in the long-wavelength limit.

For the on-site and inter-layer hopping terms, we have
\begin{equation}
\begin{aligned}
  \sum_{j}\mathbf{c}_{k_{z}j}^{\dagger}(\mathcal{H}^{k_{z}}_{\perp}+\mathcal{H}_\mathrm{ex})\mathbf{c}_{k_{z}j}= & \sum_{j}\sum_{\mathbf{\Pi}}\sum_{\mathbf{\Pi}'}\mathbf{c}_{k_{z}\mathbf{\Pi}}^{\dagger}(\mathcal{H}^{k_{z}}_{\perp}+\mathcal{H}_\mathrm{ex})\mathbf{c}_{k_{z}\mathbf{\Pi}'}e^{ i\left[(\mathcal{S} \cdot (\boldsymbol{\Pi'-\Pi}))^T \cdot \mathbf{r}_{j}\right] }  \\
= & \sum_{\mathbf{\Pi}}\sum_{\mathbf{\Pi}'}\mathbf{c}_{k_{z}\mathbf{\Pi}}^{\dagger}(\mathcal{H}^{k_{z}}_{\perp}+\mathcal{H}_\mathrm{ex})\mathbf{c}_{k_{z}\mathbf{\Pi}'}\delta(\mathcal{S} \cdot (\boldsymbol{\Pi'-\Pi})  )  \\
= & \sum_{\mathbf{\Pi}}\sum_{\mathbf{\Pi}'}\mathbf{c}_{k_{z}\mathbf{\Pi}}^{\dagger}(\mathcal{H}^{k_{z}}_{\perp}+\mathcal{H}_\mathrm{ex})\mathbf{c}_{k_{z}\mathbf{\Pi}'}\delta(\boldsymbol{\Pi'-\Pi} )  \\
= & \sum_{\mathbf{\Pi}}\mathbf{c}_{k_{z}\mathbf{\Pi}}^{\dagger}(\mathcal{H}^{k_{z}}_{\perp}+\mathcal{H}_\mathrm{ex})\mathbf{c}_{k_{z}\mathbf{\Pi}}
\end{aligned}\label{onsite-pi}
\end{equation}
Let $\mathbf{d}_{jk}=\mathbf{r}_{k}-\mathbf{r}_{j}$ denote intralayer hopping vectors. Then, the in-plane contribution becomes
\begin{equation}
\begin{aligned}
  \sum_{\langle j,\ k \rangle }\mathbf{c}_{k_{z}j}^{\dagger}(\mathcal{H}_{\parallel}+\mathcal{H}_\mathrm{wp})\mathbf{c}_{k_{z}k}
  = & \sum_{\langle j,\ k \rangle }\sum_{\mathbf{\Pi}}\sum_{\mathbf{\Pi}'}\mathbf{c}_{k_{z}\mathbf{\Pi}}^{\dagger}(\mathcal{H}_{\parallel}+\mathcal{H}_\mathrm{wp})\mathbf{c}_{k_{z}\mathbf{\Pi}'}e^{ -i\left[(\mathcal{S} \cdot \boldsymbol{\Pi})^T \cdot \mathbf{r}_{j}\right] } e^{ i\left[(\mathcal{S} \cdot \boldsymbol{\Pi'})^T \cdot \mathbf{r}_{k}\right] } \\
  = & \sum_{\langle j,\ k \rangle }\sum_{\mathbf{\Pi}}\sum_{\mathbf{\Pi}'}\mathbf{c}_{k_{z}\mathbf{\Pi}}^{\dagger}(\mathcal{H}_{\parallel}+\mathcal{H}_\mathrm{wp})\mathbf{c}_{k_{z}\mathbf{\Pi}'}  e^{ -i\left[(\mathcal{S} \cdot \boldsymbol{\Pi})^T \cdot \mathbf{r}_{j}\right] } e^{ i\left[(\mathcal{S} \cdot \boldsymbol{\Pi'})^T \cdot (\mathbf{r}_{j}+\mathbf{d}_{jk})\right] }  \\
  = & \sum_{j}\sum_{\{\mathbf{d}\}_{j}}\sum_{\mathbf{\Pi}}\sum_{\mathbf{\Pi}'}\mathbf{c}_{k_{z}\mathbf{\Pi}}^{\dagger}(\mathcal{H}_{\parallel}+\mathcal{H}_\mathrm{wp})\mathbf{c}_{k_{z}\mathbf{\Pi}'} \delta({ \mathcal{S} \cdot (\boldsymbol{\Pi-\Pi'})}) e^{ i\left[(\mathcal{S} \cdot \boldsymbol{\Pi'})^T \cdot \mathbf{d}\right] }   \\
 \approx & \sum_{\mathbf{\Pi}}\mathbf{c}_{k_{z}\mathbf{\Pi}}^{\dagger}\left(\mathcal{H}_{\parallel}(\Pi)+\mathcal{H}_\mathrm{wp}(\Pi)\right)\mathbf{c}_{k_{z}\mathbf{\Pi}},\\
\label{hopping-pi}
\end{aligned}
\end{equation}
where
\begin{equation}
\mathcal{H}_{n}(\mathbf{\Pi})=\sum_{\{\mathbf{d}\}_{j}}\mathcal{H}_{n}e^{ i\left[(\mathcal{S} \cdot \boldsymbol{\Pi'})^T \cdot \mathbf{d}\right] }\simeq\lim _{V \rightarrow \infty} \frac{1}{V} \int d \mathbf{r} \mathcal{P}(\mathbf{r}) \mathcal{H}_{n} e^{i\left[(\mathcal{S} \cdot \boldsymbol{\Pi})^T \cdot \mathbf{r}\right]}.\label{patterson}
\end{equation}
Here we have adopted the average coordination approximation, which is valid in the long-wavelength limit. In this description, $\mathcal{P}(\mathbf{r})$ is the intersite vector distribution function, which is also known as the Patterson function \cite{wolny1998average}.
In 8-fold ABT quasicrystals, the reciprocal space can be represented as integer-valued combinations of 4 2D projected vectors. The projection matrix takes the form
\begin{equation}
\mathcal{S}=\left(\begin{array}{llll}
2 \pi & 2 \pi \cos \left(\frac{ \pi}{4}\right) & 2 \pi \cos \left(\frac{ \pi}{2}\right) & 2 \pi \cos \left(\frac{3 \pi}{4}\right) \\
0 & 2 \pi \sin \left(\frac{ \pi}{4}\right) & 2 \pi \sin \left(\frac{ \pi}{2}\right) & 2 \pi \sin \left(\frac{3\pi}{4}\right)
\end{array}\right).
\end{equation}
For a reciprocal vector $\boldsymbol{\Pi}=\left(m_1, m_2, m_3, m_4\right)$ with $m_i\in \mathbb{Z}$, the corresponding Hamiltonian in momentum space is
\begin{equation}
\mathcal{H}(\mathbf{\Pi})=\mathcal{H}^{k_{z}}_{\perp}+\mathcal{H}_\mathrm{ex}+\mathcal{H}_{\parallel}+\mathcal{H}_\mathrm{wp}.
\end{equation}
Because the set $\mathcal{S}\cdot \mathbf{\Pi}$ densely fills the pseudo-Brillouin zone, we can denote $(k_{x},\ k_{y})^{T}=\mathcal{S}\cdot \mathbf{\Pi}$, where $(k_{x},\ k_{y})$ are interpreted as the physical in‑plane crystal momenta within the pseudo-Brillouin zone.

By expanding around the $\Gamma$ point, we obtain the effective $k\cdot p$ Hamiltonian for the quasicrystal. Up to second order of $\mathbf{k}$, we have
\begin{equation}
\begin{aligned}
\mathcal{H}_{\parallel} = & 2\mathcal{P}C_{1}(-4+k_{x}^{2}+k_{y}^{2})s_{0}\sigma_{0} \\
 & + 2\mathcal{P}M_{1}(-4+k_{x}^{2}+k_{y}^{2})s_{0}\sigma_{3} \\
 & -4\mathcal{P}v(k_{x}s_{1}+k_{y}s_{2})\sigma_{1}+O(\mathbf{k}^{3}),\  \\
\mathcal{H}_\mathrm{wp} = & \frac{1}{4}\mathcal{P}g\left( k_{x}^{2}k_{y}^{2}-\frac{1}{6}(k_{x}^{4}+k_{y}^{4} )\right)s_{0}\sigma_{2}+O(\mathbf{k^{5}}).
\end{aligned}
\end{equation}
where $\mathcal{P}$ is the intersite vector distribution function (Patterson function) that characterizes the effective coordination of the quasicrystal lattice.

To sum up, the total $k\cdot p$ Hamiltonian around the $\Gamma$ point is
\begin{equation}
\begin{aligned}
\mathcal{H}(\mathbf{k})=& [\widetilde{C}+\widetilde{C}_{1}k_{z}^{2}+\widetilde{C}_{2}(k_{x}^{2}+k_{y}^{2})]s_{0}\sigma_{0} \\
&+ [\widetilde{M}+\widetilde{M}_{1}k_{z}^{2}+\widetilde{M}_{2}(k_{x}^{2}+k_{y}^{2})]s_{0}\sigma_{3}\\
 & -v_{1}k_{z}s_{3}\sigma_{1}-v_{2}(k_{x}s_{1}+k_{y}s_{2})\sigma_{1} \\
 & +\widetilde{g}\left( k_{x}^{2}k_{y}^{2}-\frac{1}{6}(k_{x}^{4}+k_{y}^{4} )\right)s_{0}\sigma_{2} \\
 & +(m_xs_1+m_ys_2+m_zs_3)\sigma_{0},\\
\end{aligned}\label{Hkp_Gamma}
\end{equation}
where $\widetilde{C}=C_{0}-2C_{1}-8\mathcal{P}C_{1},$$\ \widetilde{C}_{1}=C_{1},\ \widetilde{C}_{2}=2\mathcal{P}C_{1},$$\ \widetilde{M}=M_{0}-2M_{1}-8\mathcal{P}M_{1}$$,\ \widetilde{M}_{1}=M_{1}$$,\ \widetilde{M}_{2}=2\mathcal{P}M_{1}$$,\ v_{1}=2v$$,\ v_{2}=4\mathcal{P}v$$,\ \widetilde{g}=1/4\mathcal{P}g$.

\section{Projected Surface Hamiltonian}
\label{app:B}
We now apply the standard projection method to derive the surface Hamiltonian. Starting from the $k$-space Hamiltonian of the quasicrystal in Eq.~\eqref{Hkp_Gamma}, we treat the octagonal warping term and in-plane magnetic exchange term as perturbations and focus initially on the unperturbed part of $\mathcal{H}(\mathbf{k})$.

To analyze a surface parallel to $\hat{z}$ and normal to the vector $\vec{e}_1=\cos \zeta \hat{e}_x+\sin \zeta \hat{e}_y$, we define a rotated coordinate basis for both position and momentum,
\begin{equation}
\begin{array}{ll}
x=x_1 \cos \zeta-x_2 \sin \zeta, & k_x=k_1 \cos \zeta-k_2 \sin \zeta, \\
y=x_1 \sin \zeta+x_2 \cos \zeta, & k_y=k_1 \sin \zeta+k_2 \cos \zeta.
\end{array}
\end{equation}
We define the unperturbed Hamiltonian $\mathcal{H}_\mathbb{O}$ as:
\begin{equation}
\begin{aligned}
\mathcal{H}_\mathbb{O}(\mathbf{k})= & [\widetilde{C}+\widetilde{C}_{1}k_{z}^{2}+\widetilde{C}_{2}(k_{x}^{2}+k_{y}^{2})]s_{0}\sigma_{0} \\
 & +[\widetilde{M}+\widetilde{M}_{1}k_{z}^{2}+\widetilde{M}_{2}(k_{x}^{2}+k_{y}^{2})]s_{0}\sigma_{3} \\
 & -v_{1}k_{z}s_{3}\sigma_{1}-v_{2}(k_{x}s_{1}+k_{y}s_{2})\sigma_{1} \\
 & +m_{z}s_{3}\sigma_{0}.
\end{aligned}
\end{equation}
Here, we have assumed a canted FM order with $\mathbf{m}_l = m_z \hat{z}+ m_\parallel \hat{n}$. The out-of-plane component $\mathcal{H}_\mathrm{ex} (m_z)$ is included in the unperturbed Hamiltonian, while the in-plane exchange term $\mathcal{H}_\mathrm{ex}(m_\parallel)$ and the warping term $\mathcal{H}_\mathrm{wp}$ are treated perturbatively.

In terms of the rotated coordinate $(k_{1},\ k_{2})$, the unperturbed Hamiltonian $\mathcal{H}_\mathbb{O}$ becomes
\begin{equation}
\begin{aligned}
\mathcal{H}_\mathbb{O}(\mathbf{k})=  & [\widetilde{C}+\widetilde{C}_{1}k_{z}^{2}+\widetilde{C}_{2}(k_{1}^{2}+k_{2}^{2})]s_{0}\sigma_{0}  \\
 &+ [\widetilde{M}+\widetilde{M}_{1}k_{z}^{2}+\widetilde{M}_{2}(k_{1}^{2}+k_{2}^{2})]s_{0}\sigma_{3}\\
 & -v_{1}k_{z}s_{3}\sigma_{1}-v_{2}(k_{1}s_{1}(\zeta)+k_{2}s_{2}(\zeta))\sigma_{1} \\
  & +m_{z}s_{3}\sigma_{0},
\end{aligned}
\end{equation}
where the rotated spin matrices are given by
\begin{equation}
s_{1}(\zeta)=\begin{bmatrix}
0 & e^{ -i\zeta } \\
e^{ i\zeta } & 0
\end{bmatrix},\quad s_{2}(\zeta)=\begin{bmatrix}
0 & -ie^{ -i\zeta } \\
ie^{ i\zeta } & 0
\end{bmatrix}.
\end{equation}
For a semi-infinite system defined on $x_{1}>0$, we replace $k_{1}$ by $-i\partial_{x_{1}}$. Since we are interested in low-energy physics near the $\Gamma$ point, we decompose $\mathcal{H}_\mathbb{O}$ into perpendicular and parallel parts, $\mathcal{H}_{\mathbb{O}\perp}(k_{1})$ and $\mathcal{H}_{\mathbb{O}\parallel}(k_{2},k_{z})$.
We consider a trial surface-state solution with energy $E$,
\begin{equation}
\psi=\Phi e^{-\lambda x_1}, \quad \mathcal{H}_{\mathbb{O}\perp} \psi=E \psi .
\end{equation}
To guarantee a solution, the determinant $|\mathcal{H}_{\mathbb{O}\perp}-E|$ must be equal to 0. Imposing the boundary condition $\psi(x_{1}=0)=0$, Solving these equations yields eigenvalues $E_i$ and eigenvectors composed of spinor components $\Phi_i$ and evanescent envelopes $\varphi_i$:
\begin{equation}
E_{1}=-\frac{\widetilde{C}_{2}}{\widetilde{M}_{2}}(\widetilde{M}+m_{z}), \ \psi_1=\Phi_{1}\varphi_1(x_1)=\sqrt{ \frac{\widetilde{M}_{2}+\widetilde{C}_{2}}{2\widetilde{M}_{2}} }\begin{pmatrix}
\sqrt{ \frac{\widetilde{M}_{2}-\widetilde{C}_{2}}{\widetilde{M}_{2}+\widetilde{C}_{2}} } \\
0 \\
0 \\
ie^{ i\zeta }
\end{pmatrix}A_1(e^{-\lambda_1x_1}-e^{-\lambda_2x_1}),
\end{equation}
\begin{equation}
E_{2}=\frac{\widetilde{C}_{2}}{\widetilde{M}_{2}}(-\widetilde{M}+m_{z}), \ \psi_2=\Phi_{2}\varphi_2(x_1)=\sqrt{ \frac{\widetilde{M}_{2}-\widetilde{C}_{2}}{2\widetilde{M}_{2}} }\begin{pmatrix}
0 \\
\sqrt{ \frac{\widetilde{M}_{2}+\widetilde{C}_{2}}{\widetilde{M}_{2}-\widetilde{C}_{2}} } \\
-ie^{ i\zeta  } \\
0
\end{pmatrix}A_2(e^{-\lambda_3x_1}-e^{-\lambda_4x_1}),
\end{equation}
where the normalization factors are $$A_{1}=\sqrt{\frac{2 \lambda_1 \lambda_2\left(\lambda_1+\lambda_2\right)}{\left(\lambda_1-\lambda_2\right)^2}},\; A_2=\sqrt{\frac{2 \lambda_3 \lambda_4\left(\lambda_3+\lambda_4\right)}{\left(\lambda_3-\lambda_4\right)^2}}.$$
The attenuation constants $\lambda_{1,2}$, which are determined by characteristic equations derived from boundary matching conditions, satisfy
\begin{equation}
    \lambda^{2}-\frac{\tilde{M}+m_{z}}{\tilde{M}_{2}} =\frac{v_{2}\lambda}{\sqrt{ \tilde{M}_{2}^{2}-\tilde{C}_{2}^{2} }},
\end{equation}
while $\lambda_{3,4}$ satisfy,
\begin{equation}
    \lambda^{2}-\frac{-\tilde{M}+m_{z}}{-\tilde{M}_{2}} =\frac{v_{2}\lambda}{\sqrt{ \tilde{M}_{2}^{2}-\tilde{C}_{2}^{2} }}.
\end{equation}

Projecting $\mathcal{H}_\mathbb{O}(\mathbf{k})$ onto these states and dropping the constant term, we obtain the matrix element of the unperturbed surface Hamiltonian
\begin{equation}
[\mathcal{H}_{\text {surf }}^0]_{ij}=\left\langle\psi_i\right| \mathcal{H}_\mathbb{O}(\mathbf{k})\left|\psi_j\right\rangle=\int_{0}^\infty dx\ \varphi_{i}^*\langle \Phi_{i}|\mathcal{H}_\mathbb{O}(\mathbf{k})|{\Phi_{j}} \rangle\varphi_{j}.
\end{equation}
The unperturbed surface Hamiltonian takes the form
\begin{equation}
\mathcal{H}^0_{\mathrm{surf}}=-\bar{v}_{1}k_{z}\sigma_{1}-\bar{v}_{2}k_{2}\sigma_{3}-\bar{m}\sigma_{3},
\end{equation}
where $\bar{v}_{1}=abv_{1},\ \bar{v}_{2}=av_{2}\cos 2\zeta$ with $a=\frac{\sqrt{\widetilde{M}_{2}^{2}-\widetilde{C}_{2}^{2}  }}{\widetilde{M}_{2}}$, $b=\langle \varphi_1|\varphi_2\rangle=A_{1}A_{2}\left(\frac{1}{\lambda_1 + \lambda_3} - \frac{1}{\lambda_2 + \lambda_3} - \frac{1}{\lambda_1 + \lambda_4} + \frac{1}{\lambda_2 + \lambda_4}\right)$, and $\bar{m}=\frac{m_{z}\widetilde{C}_{2}}{\widetilde{M}_{2}}$.

Next, we project the perturbation term $\mathcal{H}_\mathbb{I}=\mathcal{H}_\mathrm{ex}(m_\parallel)+\mathcal{H}_\mathrm{wp}$ onto these surface states.
For the warping term $\mathcal{H}_\mathrm{wp}$, substituting $k_1$ by $-i\partial_{x_1}$ gives
\begin{equation}
\begin{aligned}
   \mathcal{H}_{\mathrm{wp}}= \tilde{g}\left(-\partial_{x_1}^2 k_2^2-\frac{1}{6}\left(\partial_{x_1}^4+k_2^4\right)\right)\cos(4\zeta) s_0 \sigma_2.
\end{aligned}
\end{equation}
The matrix elements are
\begin{equation}
    \begin{aligned}
        [\mathcal{H}_{\text {surf }}^{\mathrm{wp}}]_{ij}&=\left\langle\psi_i\right| \mathcal{H}_{\mathrm{wp}}\left|\psi_j\right\rangle=\int_0^{\infty} dx \varphi_i^*\left\langle\Phi_i\right| \mathcal{H}_{\mathrm{wp}}\left(-i \partial_{x_1}\right)\left|\Phi_j\right\rangle \varphi_j \\
&= \tilde{g}\cos(4\zeta)\langle{\Phi_{i}} |s_{0}\sigma_{2}|{\Phi_{j}} \rangle\int_0^{\infty}dx\varphi_{i}^* \left(-\partial_{x_1}^2 k_2^2-\frac{1}{6}\left(\partial_{x_1}^4+k_2^4\right)\right)\varphi_{j}.
    \end{aligned}
\end{equation}
Since $\langle{\Phi_{i}} |s_{0}\sigma_{2}|{\Phi_{j}} \rangle=[a\sigma_{2}]_{ij}$, we define
\begin{equation}
    \int_0^{\infty}\langle  {\varphi_{1}} |\left(-\partial_{x_1}^2 k_2^2-\frac{1}{6}\left(\partial_{x_1}^4+k_2^4\right)\right)|{\varphi_{2}}  \rangle=-\frac{1}{6}bk_{2}^4-\frac{1}{6}\kappa_{4}-\kappa_{2}k_{2}^2,
\end{equation}
with
\begin{equation}
\kappa_{4}=\int_0^{\infty}\left\langle   {\varphi_{1}} |\partial_{x_1}^4|{\varphi_{2}}   \right\rangle=\left( \lambda_{3}^{4}\left( \frac{1}{\lambda_{1}+\lambda_{3}}-\frac{1}{\lambda_{2}+\lambda_{3}}\right) -\lambda_{4}^{4}\left( \frac{1}{\lambda_{2}+\lambda_{4}}-\frac{1}{\lambda_{1}+\lambda_{4}} \right) \right),
\end{equation}
\begin{equation}
\kappa_{2}=\int_0^{\infty}\left\langle   {\varphi_{1}} |\partial_{x_1}^2|{\varphi_{2}}   \right\rangle=\frac{(\lambda_{1}-\lambda_{2})(\lambda_{3}-\lambda_{4})(\lambda_{1}\lambda_{2}\lambda_{3}+\lambda_{2}\lambda_{3}\lambda_{4}+\lambda_{1}\lambda_{2}\lambda_{4}+\lambda_{1}\lambda_{3}\lambda_{4})}{(\lambda_{1}+\lambda_{3})(\lambda_{2}+\lambda_{3})(\lambda_{1}+\lambda_{4})(\lambda_{2}+\lambda_{4})}.
\end{equation}

Defining $$\mathcal{M}^{\mathrm{wp}}(k_{2})=(-\frac{1}{6}bk_{2}^4-\frac{1}{6}\kappa_{4}-\kappa_{2}k_{2}^2)a\tilde{g},$$ we obtain
\begin{equation}
    \mathcal{H}_{\text {surf }}^{\mathrm{wp}}=\mathcal{M}^{\mathrm{wp}}(k_{2}) \cos (4 \zeta) \sigma_2 .
\end{equation}
Since the Dirac cone of $\mathcal{H}^{0}_{\mathrm{surf}}$ is located at $k_{2}=-\bar{m}/v_2$, the warping term induces a mass term
\begin{equation}
    \mathcal{H}^\mathrm{wp}_{\mathrm{surf}}=\mathcal{M}^{\mathrm{wp}}_0 \cos (4 \zeta) \sigma_2 ,\quad \mathcal{M}_0^{\text {wp }}=\mathcal{M}^{\mathrm{wp}}(-\frac{\bar{m}}{v_{2}})=\left(-\frac{1}{6}b(\frac{\bar{m}}{v_{2}})^4-\kappa_{2}(\frac{\bar{m}}{v_{2}})^2-\frac{1}{6}\kappa_{4}\right)a\tilde{g}.
\end{equation}

Finally, projecting the in-plane exchange term yields
\begin{equation}
[\mathcal{H}_{\mathrm{surf }}^{\mathrm {ex}}]_{ij}=\left\langle\psi_i\right| \mathcal{H}_{\mathrm {ex}}\left|\psi_j\right\rangle.
\end{equation}
Denoting $m_{\|}=m \sin \theta$ with $m_x=m_{\|} \cos \phi, m_y=m_{\|} \sin \phi$, we obtain
\begin{equation}
\begin{aligned}
\mathcal{H}^\mathrm{ex}_{\mathrm{surf}}
= (m_{x}\cos \zeta+m_{y}\sin \zeta)ab\sigma_{2}=m_{\parallel}ab\cos (\zeta-\phi)\sigma_{2}.
\end{aligned}
\end{equation}

\end{widetext}

\providecommand{\noopsort}[1]{}\providecommand{\singleletter}[1]{#1}%

\end{document}